\documentclass[12pt,preprint]{aastex}

\shorttitle{A magnetic reconnection source of anomalous cosmic rays}
\shortauthors{Drake et al.}

\begin{document}


\title{Dissipation of the sectored heliospheric magnetic field near the heliopause: a mechanism for the generation of anomalous cosmic rays}


\author{J.~F.~Drake\altaffilmark{1}, M.~Opher\altaffilmark{2}, M.~Swisdak\altaffilmark{1} and J.~N.~Chamoun\altaffilmark{1}}


\altaffiltext{1}{University of Maryland, College Park, MD 20742;
drake@umd.edu, swisdak@umd.edu}

\altaffiltext{2}{George Mason University, 4400 University Drive, Fairfax, VA 22030; mopher@gmu.edu}


\begin{abstract}
The recent observations of the anomalous cosmic ray (ACR) energy
spectrum as Voyagers 1 and 2 crossed the heliospheric termination
shock have called into question the conventional shock source of these
energetic particles. We suggest that the sectored heliospheric
magnetic field, which results from the flapping of the heliospheric
current sheet, piles up as it approaches the heliopause, narrowing the
current sheets that separate the sectors and triggering the onset of
collisionless magnetic reconnection. Particle-in-cell simulations
reveal that most of the magnetic energy is released and most of this
energy goes into energetic ions with significant but smaller amounts
of energy going into electrons. The energy gain of the most energetic
ions results from their reflection from the ends of contracting
magnetic islands, a first order Fermi process. The energy gain of the ions in
contracting islands increases their parallel (to the magnetic field
${\bf B}$) pressure $p_\parallel$ until the marginal firehose
condition is reached, causing magnetic reconnection and associated
particle acceleration to shut down. Thus, the feedback of the
self-consistent development of the energetic ion pressure on
reconnection is a crucial element of any reconnection-based,
particle-acceleration model. The model calls into question the strong
scattering assumption used to derive the Parker transport equation and
therefore the absence of first order Fermi acceleration in
incompressible flows. A simple 1-D model for particle energy gain and
loss is presented in which the feedback of the energetic particles on
the reconnection drive is included. The ACR differential energy
spectrum takes the form of a power law with a spectral index slightly
above $1.5$. The model has the potential to explain several key
Voyager observations, including the similarities in the spectra of
different ion species.

\end{abstract}





\section{INTRODUCTION}
\label{intro}
Anomalous Cosmic Rays (ACRs) are ions that have energies in the range
of $5-100MeV/$nucleon, just below energies associated with galactic
cosmic rays. It is known that the ACRs are produced from interstellar
neutral atoms since their composition matches that of the local
interstellar medium (LISM) \citep{Cummings96,Cummings07}. In the standard model
the LISM neutrals are ionized and picked up by the solar wind deep
within the heliosphere and carried out to the heliospheric
termination shock, where they undergo diffusive shock acceleration
\citep{Pesses81}. The termination shock (TS) marks the transition of
the solar wind from supersonic to subsonic flow.

A major surprise when the Voyager spacecraft crossed the TS was that
the number density of the ACRs did not peak at the shock
\citep{Stone05,Stone08}, indicating that their source was
elsewhere. Since Voyager 1's crossing of the TS in December 2004 and
Voyager 2's crossing in August-September 2007, the two spacecraft have
seen increasing intensities of ACRs as they penetrate further into the
heliosheath (HS), the subsonic plasma downstream from the shock. One
possible explanation for the Voyager observations is that the ACRs
peak along the flanks of the heliosphere where the spiral magnetic
field from the sun has been in contact with the TS for a longer period
of time \citep{McComas06}. A compressional model of ACRs has also been
proposed \citep{Fisk06a}.

Here we suggest that it is not the TS that is the source of the
ACRs. Instead it is the annihilation of the ``sectored'' magnetic
field within the heliosheath as it is compressed on its approach to
the heliopause (HP) that produces the ACRs. The dipole magnetic
field emanating from the sun is quickly dragged in the azimuthal
direction due to the sun's rotation. The resulting azimuthal field
$B_\phi$ reverses sign across the equator with the two polarities
separated by the heliospheric current sheet. The difference between
the rotation axis of the sun and the magnetic pole of the sun causes
the heliospheric current sheet to flap in the vertical direction with
a period of $26$ days. The resulting folded heliospheric current sheet
creates a heliospheric field that displays a ``sector'' structure
consisting of alternating signs of $B_\phi$ \citep{Wilcox65}. The
``sector-zone'' occupies a latitudinal extent that varies during the
solar cycle, reaching nearly the poles when the fields from the sun
are a maximum \citep{Smith01}. The sectors remain a prominent feature
of the heliospheric magnetic field out to the TS although the period
of the reversals varies erratically from the nominal $13$ day value
\citep{Burlaga03,Burlaga05,Burlaga06}. Within the heliosheath the
period of the reversals is even more variable, probably a consequence
of the low flow speeds in this region and the time-varying location of
the TS \citep{Burlaga06}.

An obvious question is: Why does the sectored azimuthal field survive out
to distances of order $90$AU? Why don't the reversed fields simply
annihilate as a result of magnetic reconnection, releasing the stored
magnetic energy? The heliosphere is, for all practical purposes,
collisionless. It is well known that the rate of collisionless
reconnection drops dramatically when the width of the current channel
is greater than the ion inertial scale $d_i=c/\omega_{pi}$ where
$\omega_{pi}$ is the ion plasma frequency
\citep{Yamada07,Cassak05}. At $1$AU the current sheet is on average
around $10,000$km wide, which is in the range of $200d_i$
\citep{Winterhalter94,Smith01}. The deviations from this mean width are
substantial and magnetic reconnection is occasionally observed during
crossings of the heliospheric current sheet \citep{Gosling07a}. In any
case, upstream of the TS the heliospheric current sheet is too wide to
expect collisionless reconnection to dissipate the sector zone.

In the present paper we present MHD simulations of the global
heliosphere showing the development of the sectored heliospheric
field. The current sheet compresses and the spacing of the current
sheets shrinks across the TS. As the plasma in the heliosheath
approaches the heliopause its radial motion slows, causing the current
sheet widths and sector spacing to further decrease. Current sheets
upstream of the Earth's bow shock have been observed to compress by
more than a factor of $30$ as they first compress across the shock and
then pile up against the magnetopause \citep{Phan07}. At the same time
the magnetic field strength rises and the plasma density and pressure
decrease as the plasma is squeezed away from the nose of the
heliosphere along the magnetic field.  The resulting drop in the local
plasma $\beta$, the ratio of plasma to magnetic pressure, is similar
to that seen just upstream of the Earth's magnetopause
\citep{Crooker79}.

When the width of the heliospheric current layers separating the
sectored fields approach the ion inertial scale, the sectored field
undergoes magnetic reconnection. In particle-in-cell (PIC) simulations
of the sectored geometry, we demonstrate that essentially all of the
magnetic energy in the ``sector zone'' is released into energetic
particles as the sectors form a complex network of interacting
magnetic islands. 

A number of mechanisms have been proposed to explain ion acceleration
during reconnection, mostly to address flare observations
\citep{Miller97}. Parallel electric fields are more important for
electrons than ions \citep{Litvinenko96,Pritchett08} and the
localization of parallel electric fields in any case limits their
overall importance \citep{Drake06,Egedal09}. The resonance absorption
of cascading MHD turbulence resulting from reconnection has also been
proposed as an accelerator of ions \citep{Miller98} although
the mechanism for the efficient generation of this turbulence remains
unspecified. The Petschek slow shocks bounding the reconnection
outflow have also been proposed as sites of ion acceleration
\citep{Tsuneta96}. However, satellite crossings of these boundary
layers indicate that ions are heated upon crossing into the
reconnection outflows but do not reveal a significant energetic
particle component \citep{Gosling05}.

We find that the most energetic ions are accelerated through
reflection in contracting islands, a first-order Fermi process that
was studied earlier for electron acceleration \citep{Drake06}. The
rate of particle energy gain is given by
\begin{equation}
\frac{dE}{dt}=\alpha\langle\frac{c_A}{L}\rangle E,
\label{rate}
\end{equation}
where $\langle c_A/L\rangle$ is a measure of the rate of contraction
of magnetic islands and $\alpha$ is a constant. Importantly, the rate
of energy gain is proportional to the particle energy and is
independent of the particle mass. The pickup particles gain the
most energy and form the ACR spectrum because of their high seed
energy. Equation (\ref{rate}) implies that the ACR spectra of ions
with different masses are similar because their seed energy and rate
of energy gain are identical when expressed on a per nucleon
basis. The energy spectra of all species assume a power law with an
index of $(3+\beta_0)/2$, where $\beta_0$ is the initial pickup ion
$\beta$ where reconnection onsets. An ACR acceleration model based on
the MHD description of turbulent reconnection of the sectored fields has also
been recently proposed \citep{Lazarian09} and its relationship with the present work is discussed in Sec.~\ref{discussion}.

\section{MHD simulations of the sector structure of the heliospheric current sheet}
\label{mhdsimulations}

To gain a better understanding of the profiles of the plasma parameters
and the sector structure of the magnetic field in the crucial region
between the TS and the HP, we have carried out MHD simulations
including the tilt in the solar magnetic field with respect to the
rotation axis. 

Our 3D MHD model includes the major components of the interaction of
the solar wind with the interstellar medium: the interplanetary and
interstellar magnetic field and the ionized and neutral H atoms that
interact through charge exchange. Although the neutral H atoms have a
mean free path on the order of $100AU$ and should be treated
kinetically, a multi-fluid description (four fluids describing each
zone of the interaction) does a comparable job (see {\it
e.g.}\citep{Alexashov05}).  There are a total of five hydrogen fluids,
one ionized and four neutral populations \citep{Opher09} in the model,
which is similar to earlier models \citep{Alexashov05,Zank96} and was
benchmarked with a kinetic model \citep{Izmodenov09}. The four
populations of neutral H atoms are evolved throughout the simulation
domain. They represent the particles of interstellar origin, those
peaked in the regions between the bow shock and heliopause, the
supersonic solar wind, and the compressed region between the
termination shock and the heliopause. All four populations are
described by separate systems of the Euler equations with
corresponding source terms \citep{McNutt99}. The ionized component
interacts with the H neutrals via charge exchange \citep{Izmodenov09}.
 
The parameters for the inner boundary (at $30AU$) are those used by
\citep{Izmodenov09} to match the observations: a proton density of
$8.7\times 10^{-3}cm-3$, a temperature of $1.087\times 10^5 K$ and a
Parker spiral field of $2nT$ at the equator.  The parameters for the
density, velocity and temperature for the ions and neutrals at the
outer boundary in the interstellar medium reflect the best
observational values: a proton density of $0.06cm^{-3}$, a velocity of
$26.3km/s$, a temperature of $6519K$, a density of neutral H in the
LISM of $0.18cm^{-3}$ with the same velocity and temperature as the
LISM protons. The only parameter with significant uncertainty at the
outer boundary is the interstellar magnetic field. Both the intensity
and the directions have large uncertainties \citep{Opher09}. We assume
an interstellar magnetic field of $4.4\mu G$ with tilt angles $\beta=
60^\circ$ (the angle between the interstellar magnetic field BISM and
the heliospheric equatorial plane) and $\alpha=20^\circ$ (the angle
between BISM and the interstellar wind). These values reproduce the
positions of the crossings of the TS by Voyagers 1 and 2
\citep{Stone05,Stone08}.

We tilted the Parker spiral at the inner boundary by $7^\circ$ with
respect to the rotation axis of the sun.  A refined grid was designed
to resolve the reversals of the heliospheric current sheet. The grid
has an inner boundary of $30AU$ and an outer boundary ranging from
$-300AU$ to $300AU$ in x, y, and z directions. The computational cell
size ranges from $0.07AU$ to $18.75AU$. We used fixed inner boundary
conditions for the ionized fluid and soft boundaries for the neutral
fluids. The outer boundaries were all outflows with the exception of
the -x boundary, where inflow conditions were imposed for the ionized
plasma and the neutrals coming from the LISM. In order to bring the TS
closer to the inner boundary to better resolve the sector
structure in the HS, we artificially lowered the solar wind speed to
$300km/s$.

\begin{figure}
\epsscale{.6}
\plotone{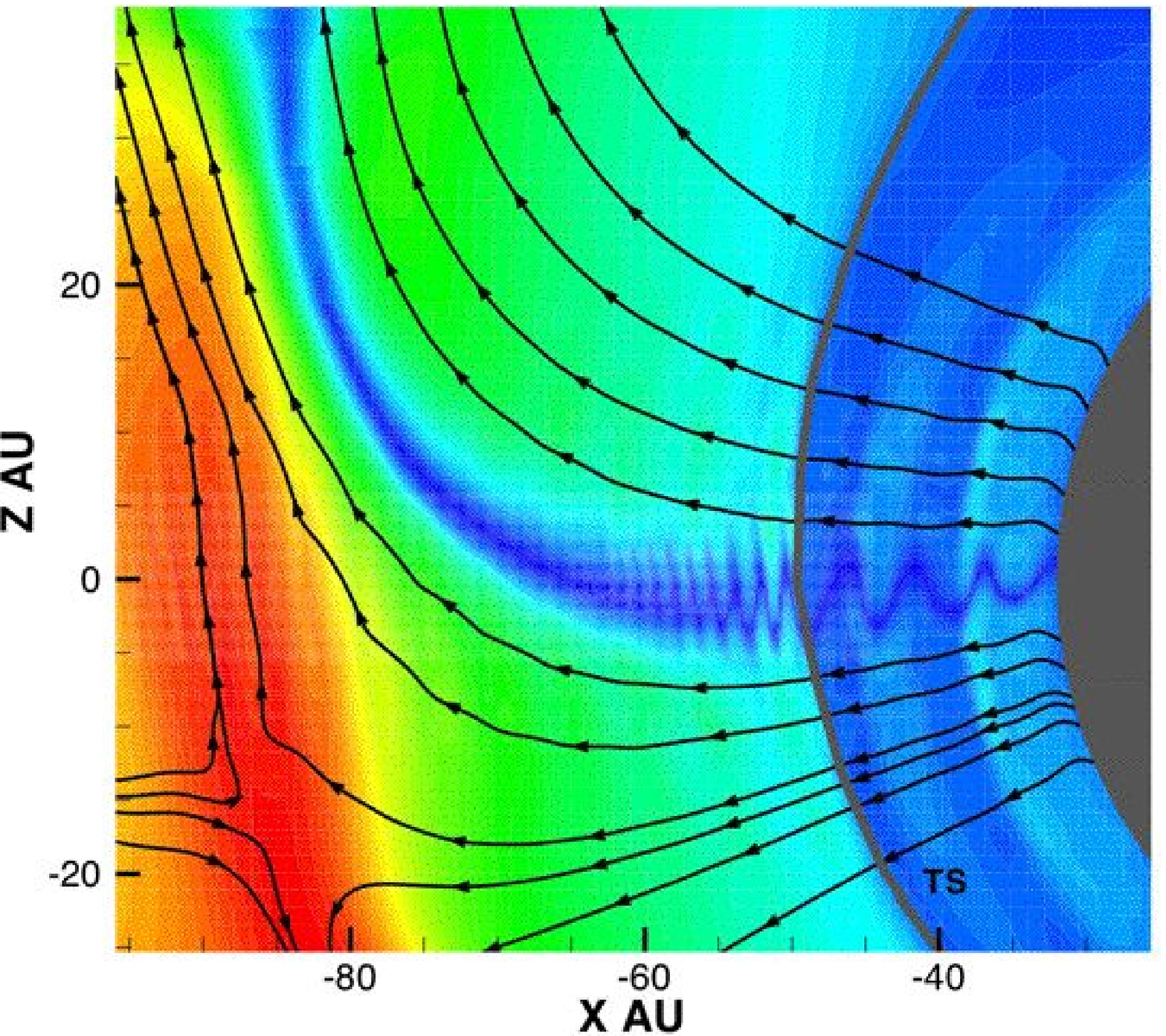}
\plotone{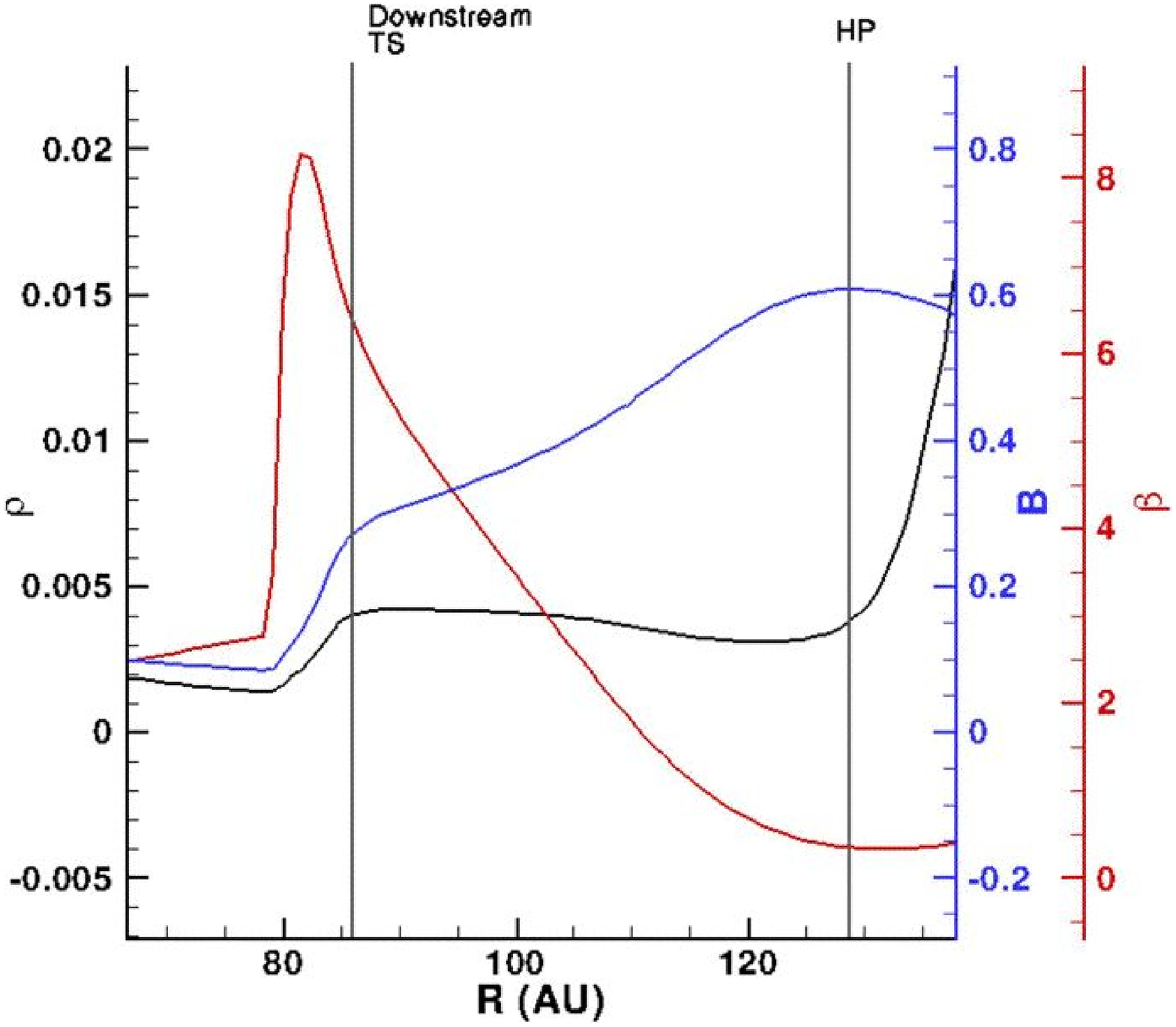}
\caption{\label{mhdB} (Color online) In (a) the heliospheric magnetic
field $B$ and the streamlines of the flow in a cut in the $x-z$ plane
from a 3-D MHD simulation with an angle between the rotation axes of
the sun and its magnetic field of $7^\circ$. The high resolution is
this simulation required that the solar wind velocity be reduced to
$300km/s$ to reduce the overall size of the heliosphere. In (b) plots
of the magnitude of $B$ in $nT$, the density $\rho$ in $cm^{-3}$ and
$\beta$ along the stagnation flowline from a simulation without a
tilted magnetic field and a solar wind velocity of $417km/s$.}
\end{figure}

In Fig.~\ref{mhdB}(a) we show the total magnetic field $B$ in the
$x-z$ plane (color) with the flow stream lines in black. Because of
the low solar wind speed the TS moves inward to $50AU$. The
oscillating region of low $B$ marks the location of the heliospheric
current sheet. Above (below) the current sheet the azimuthal field
$B_\phi$ is positive (negative). The region of sectored magnetic field
falls within the oscillations of the current sheet. We emphasize that
in spite of the high resolution used in this simulation, the
calculation significantly underestimates the oscillation amplitude of
the current sheet, which on the basis of observations is expected to
occupy a region equatorward of a fixed latitudinal angle that depends
modestly on the phase of the solar cycle. At the crossing of the TS by
Voyagers 1 and 2 the latitudinal extent of the sector region was $30$
degrees and at solar maximum the current sheet can approach the
poles. The sector structure is compressed downstream of the TS (the
sector spacing decreases from $4.7AU$ to $1.9AU$) and the spacing of
the sectors continuously decreases further into the HS as the radial
plasma velocity decreases.

We have also carried out a simulation with no relative tilt of the
rotation axes of the sun and magnetic field. This allowed reduced
resolution and a much larger simulation domain so that the solar wind
velocity could be increased to $417km/s$ to match the observed
values. In this simulation we used the same interstellar magnetic
field strength and orientation as in the simulation shown in
Fig.~\ref{mhdB}(a) For this case the heliosphere expands to a size
that closely matches the Voyager observations of the location of the
TS.  In Fig.~\ref{mhdB}(b) the variation of the plasma density $\rho$,
the magnetic field $B$ and the local plasma $\beta$ are shown along
the stagnation line, similar to that which can be seen near the bottom
edge of Fig.~\ref{mhdB}(a) (see also \citep{Opher09}). As the radial
velocity decreases as the plasma approaches the stagnation point at
the HP, the magnetic field compresses and the plasma squeezes out
along the magnetic field causing the density and pressure to
drop. These trends cause $\beta$ to fall below unity.

We emphasize that although the sectored field
region in the simulation in Fig.~\ref{mhdB}(a) does not overlap the
stagnation line (because of the $7^\circ$ tilt of the magnetic field
in the simulation), the Voyager 1 and 2 satellites bracketed the
stagnation line at the time of the TS crossing (the tilt at the time
was around $30^\circ$) and both measure the sectored magnetic field
\citep{Burlaga03}. Thus, the sectored magnetic field becomes
increasingly compressed on its approach to the HP and locally
dominates the pressure, in sharp contrast with the upstream HS.

Magnetic reconnection is expected to sharply onset in the region close
to the HP due to the compression of the heliospheric current
sheet in this region. In this low $\beta$ region the available
magnetic free energy is sufficient to accelerate the pickup ions to
the $10-100MeV/nuc$ range of energies that make up the ACRs. Since
this is a crucial result for the energetics of the model to work out,
we have checked that the result is not sensitive to the specific
boundary conditions used in the simulations shown in
Fig.~\ref{mhdB}. A similar plot of the parameters along the stagnation
of line from the simulation of Fig.~\ref{mhdB}(a) revealed that, while
the spatial separation between the TS and HP was smaller, the increase
in magnetic field strength and the minimum value of $\beta$ were
essentially identical. Thus, neither the magnitude of the tilt between
the magnetic and solar rotation axes nor the solar wind speed alter
these crucial results of the MHD simulations. Finally, since the
orientation of the interstellar magnetic field also remains uncertain,
we carried out simulations to check the sensitivity of the compression
of the magnetic field and drop in $\beta$ to these parameters. The
results were essentially identical to those shown in
Fig.~\ref{mhdB}(b).

\section{PIC simulations of reconnection and particle acceleration}
\label{picsims}

The MHD model does not reliably describe magnetic reconnection or
particle acceleration. We explore both by carrying out 2-D
particle-in-cell (PIC) simulations of the sector structure in the
heliospheric $x-y$ plane. The simulations are performed with the PIC
code p3d \citep{Zeiler02} using a periodic equilibrium magnetic field
$B_y(x)$ resulting from a series of Harris current sheets of peak
density $n_0$ superimposed on a uniform background of density
$n_b=0.2n_0$. The results are presented in normalized units: the
magnetic field to the asymptotic value of the reversed field $B_{0y}$,
the density to $n_0$, velocities to the proton Alfv\'en speed
$c_A=B_{0y}/\sqrt{4\pi m_pn_0}$, times to the inverse proton
cyclotron frequency in $B_{0y}$, $\Omega_p^{-1}=m_pc/eB_{0y}$,
lengths to the proton inertial length $d_p=c_A/\Omega_p$ and
temperatures to $m_pc_A^2$. We define some important scale lengths as
follows: $w_0=0.5d_p$ is the half-width of an individual current
sheet; $L_p=25.6d_p$ is the separation of neighboring current sheets;
$\Delta_x=\Delta_y=0.05d_p$ are the grid scales; and $L_x=204.8d_p$ and
$L_y=409.6d_p$ are the lengths of the overall computational domain.
Since we are focusing on the ion dynamics, which requires as large a
simulation domain as possible, we minimize the separation of scales by
using a modest ion to electron mass ratio of $25$ and velocity of
light $c$ of $15c_A$. The particle temperatures are initially uniform
with $T_p=(5/12)m_pc_A^2$ and $T_e=(1/12)m_pc_A^2$. A second set
of simulations have been carried out with a small number ($5\%$) of
ions with high temperature ($T_p=12.5m_pc_A^2$) to represent the
seed population of ACRs. The average number of particles per cell
is $100$ outside of the current sheets and the total number of protons
plus electrons exceeds eight billon. Reconnection begins from particle
noise.

The overall scale sizes of our simulations are much smaller than those
of the heliospheric sectored field. The widths of the sectors upstream
of the TS are around $1.7\times 10^8km$, which at a density of
$0.001/cm^3$, is around $2\times 10^4c/\omega_{pi}$. A factor of $40$
compression through the shock and the approach to the heliopause
brings the spacing down to $4\times 10^6km$, or with a density of
$0.003/cm^3$, around $1000c/\omega_{pi}$. This sector spacing is
clearly much larger than the value of $25.6c/\omega_{pi}$ in our
simulations. The value of $25.6c/\omega_{pi}$ was chosen as the
minimum value necessary to fully magnetize the ions in the simulations
\citep{Mandt94}, which is a necessary requirement to correctly assess
the mechanism for ion acceleration. The simulations reveal, as
expected, that islands on adjacent current layers eventually
overlap. At this time the rate of magnetic energy dissipation is a
maximum and the characteristic width of islands is comparable to the
sector spacing. The $25.6c/\omega_{pi}$ size islands are large enough
to magnetize ions. We also emphasize that the artificial sector
spacing used in the simulations does not affect the ultimate fate of
the sectors, that the islands grow to a large enough size to destroy
the sector structure. As long as the physical system in the direction
along the current layers (the azimuthal direction in the heliosphere)
is much longer than the sector spacing, once reconnection onsets the
sectors will completely break up into a bath of merging islands. In
our simulations the ratio $L_y/L_p=16$ is more than sufficient to
guarantee that the sectors fully reconnect. 

The 2-D assumption is also a limitation of the present model. However,
3-D models of anti-parallel reconnection reveal that the electron
current perpendicular to the plane of reconnection causes the x-lines
to lengthen so that at late time reconnection becomes quasi two-dimensional
\citep{Hesse01,Huba02,Shay03}. In contrast, in the presence of a guide
field such as in the corona, 3-D magnetic reconnection differs greatly
from that in 2-D because magnetic islands with differing pitch can
grow in 3-D but not in 2-D \citep{Onofri06,Drake06}.

Reconnection of the current sheet is very sensitive to the initial
current layer width $w_0$. The growth of islands when $w_0\gg
c/\omega_{pi}$ is strongly inhibited, a result that is consistent with
the survival of the sectored field during its propagation to the outer
heliosphere.  The observed width of the heliospheric current sheet
exhibits large variations \citep{Smith01}. Within the heliosheath
crossing times of the current sheet ranging from a day to $50s$, the
time resolution of the Voyager magnetometer, have been documented
\citep{Burlaga06}. Deducing the physical width of the current sheet
from these crossing times is difficult because of the uncertainty of
the relative velocity of the spacecraft and the ambient plasma. For the
$100km/s$ flows expected just downstream of the TS, the $50s$ crossing
translates to around $1c/\omega_{pi}$. Further compression of the
current layers on their approach to the HP similar to that measured at
the magnetopause \citep{Phan07} would further reduce $w_0$. Thus, the
value of $w_0=0.5c/\omega_{pi}$ taken in our simulations is
reasonable. In any case since there is no obvious process to halt the
compression of the current sheets as they approach the heliopause, it
seems inevitable that the current layers will compress until
reconnection onsets.

\begin{figure}
\epsscale{.6}
\plotone{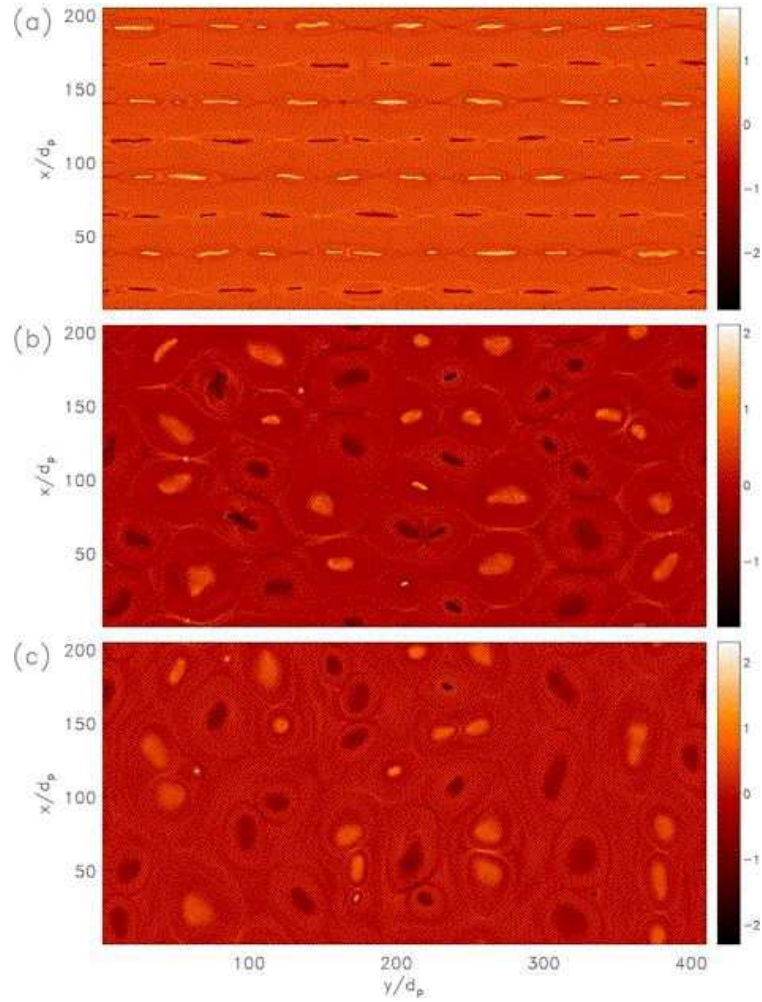}
\caption{\label{jez3t} (Color online) The out-of-plane current $J_{ez}$ in the $x-y$ plane at three times from a PIC simulation.}
\end{figure}

In Fig.~\ref{jez3t} we show the distribution of the out-of-plane
electron current $J_{ez}$ at three times. This simulation includes the
ACR seed particles. In (a) at $\Omega_pt=76$ the islands are growing
on each of the current layers. At this point in time, because of their
relatively short wavelengths, the islands on each current layer grow
independently of those on adjacent current layers. This is evident
from the lack of phase correlation between islands and x-lines on
adjacent current layers in (a). In the double-tearing mode
\citep{Pritchett80} magnetic islands on adjacent current layers line
up so that an island on one current layer drives the x-line on the
adjacent layer.  Also in (a) some of these islands are becoming larger
by coalescing with their neighbors. The merging process releases
substantial energy, essentially all of the energy in the smaller of
the merging pairs \citep{Fermo09}. In (b) at $\Omega_pt=136$ the
islands on adjacent current layers have overlapped and islands between
the same and adjacent layers are now merging. Finally in (c) at
$\Omega_pt=176$ merging continues and there is essentially no evidence
of the original sectored magnetic field. At the end of the simulation
at $\Omega_pt=200$, $72\%$ of the magnetic energy has been released,
$71\%$ going into ions and $29\%$ into electrons. The implication of
these results is that the sectored magnetic field will essentially be
entirely dissipated as it approaches the HP.

\begin{figure}
\epsscale{.6}
\plotone{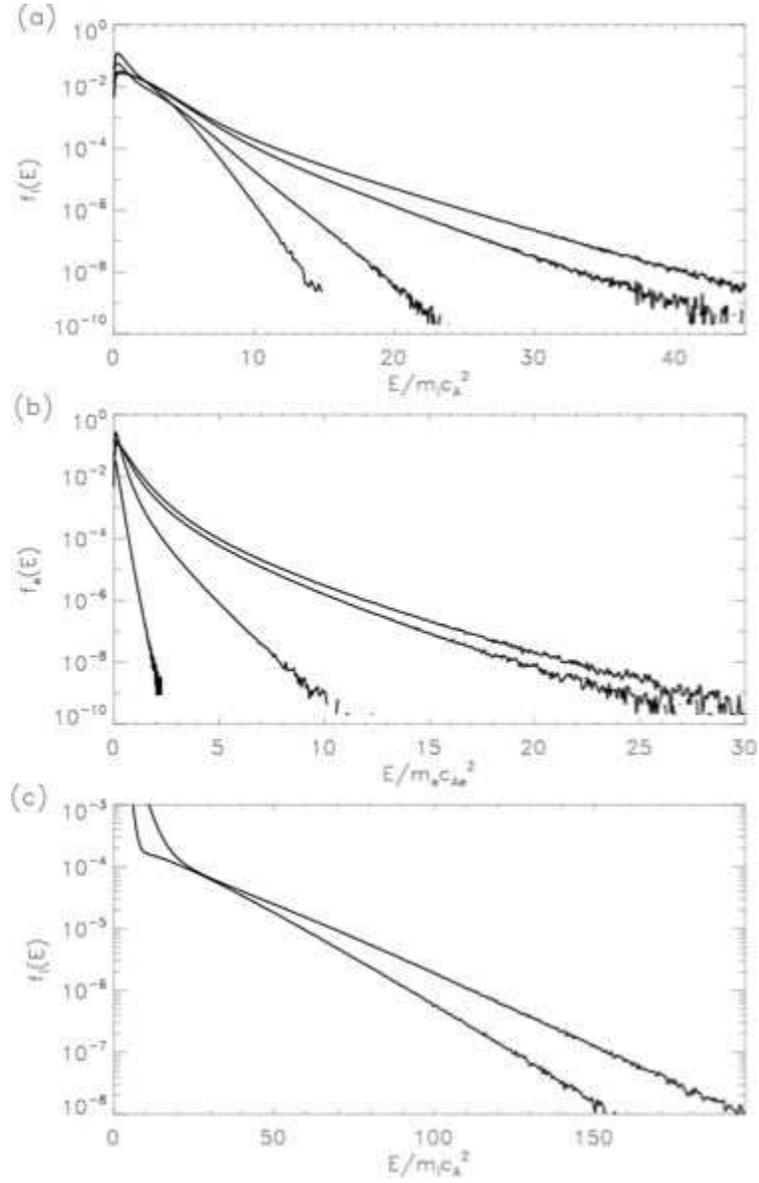}
\caption{\label{energy_eiacr} The energy spectra of (a) ions and (b)
electrons at $\Omega_pt=0$, $50$, $100$ and $200$ from a PIC
simulation without a seed population of pickup ions. In (c) the
spectra of ions with a seed poplulation of pickup ions at
$\Omega_pt=0$, $200$.}
\end{figure}

In Fig.~\ref{energy_eiacr}(a),(b) we show the energy spectra of ions
and electrons at times $\Omega_pt=0$, $50$, $100$ and $200$. This data
is from a simulation without seed ACRs so the acceleration of the core
ions can be more clearly seen. High energy tails form on both the
electron and ion energy distributions. Most of the energy gain occurs
between $50\Omega_p^{-1}$ and $100\Omega_p^{-1}$ when islands between
adjacent current layers are reconnecting. Note, however, that the
distributions are not of the form of a power law. On the other hand,
power laws are not expected since they typically arise when processes
leading to energy gain of particles are balanced by loss
\citep{Blandford87,Drake06}. We note that the energy spectra of
Fig.~\ref{energy_eiacr} are not sensitive to the sector spacing. The
electron and ion energy spectra from simulations with a sector spacing
half of that of Fig.~\ref{energy_eiacr} are nearly identical.

\begin{figure}
\epsscale{.6}
\plotone{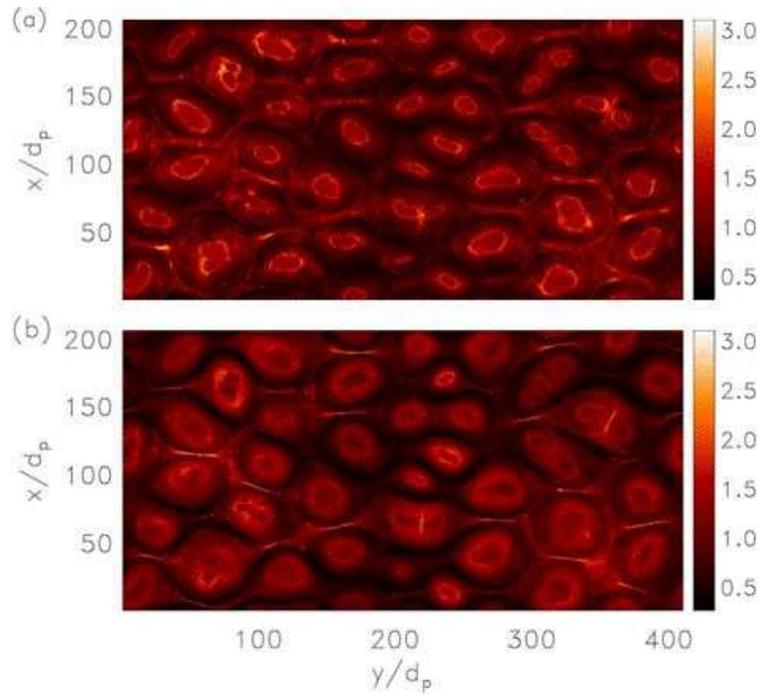}
\caption{\label{iontemp} (Color online) The ion temperature in the $x-y$ plane
(a) parallel and (b) perpendicular to the local magnetic
field at $\Omega_pt=136$ from the simulation in Fig.~\ref{jez3t}.}
\end{figure}

To gain a better understanding of the mechanism for particle
acceleration, and in particular that of the ions, we show in
Fig.~\ref{iontemp} the spatial distribution of the ion temperature
parallel (in (a)) and perpendicular (in (b)) to the local magnetic
field at $\Omega_pt=136$ from the simulation of Fig.~\ref{jez3t}. Note
that the color tables of the two plots are identical so the relative
values of $T_\parallel$ and $T_\perp$ can be seen. Strong increases in
the temperature are evident both in reconnecting current layers and
within magnetic islands. In the current layers $T_\perp$ typically
exceeds $T_\parallel$. Within magnetic islands $T_\parallel$ is
largest and forms distinct high temperature rings. One source of the
hot ions within islands is the ejecta of heated ions from the current
layers. However, the hot rings within islands appear disconnected from
current layers. The enhancement of $T_\parallel$ compared with
$T_\perp$ suggests that the ion contraction mechanism that was
proposed earlier as a mechanism for heating electrons \citep{Drake06}
is also active for ions. This mechanism requires that particles bounce
several times within an island during its contraction and therefore
that the particle velocity be super-Alfv\'enic. Evidently this is
easier for electrons than for ions. It is evident from
Fig.~\ref{energy_eiacr} that the fraction of super-Alfv\'enic ions is
signficant and the results from Fig.~\ref{iontemp} are consistent with
the contraction mechanism.

\begin{figure}
\epsscale{.6}
\plotone{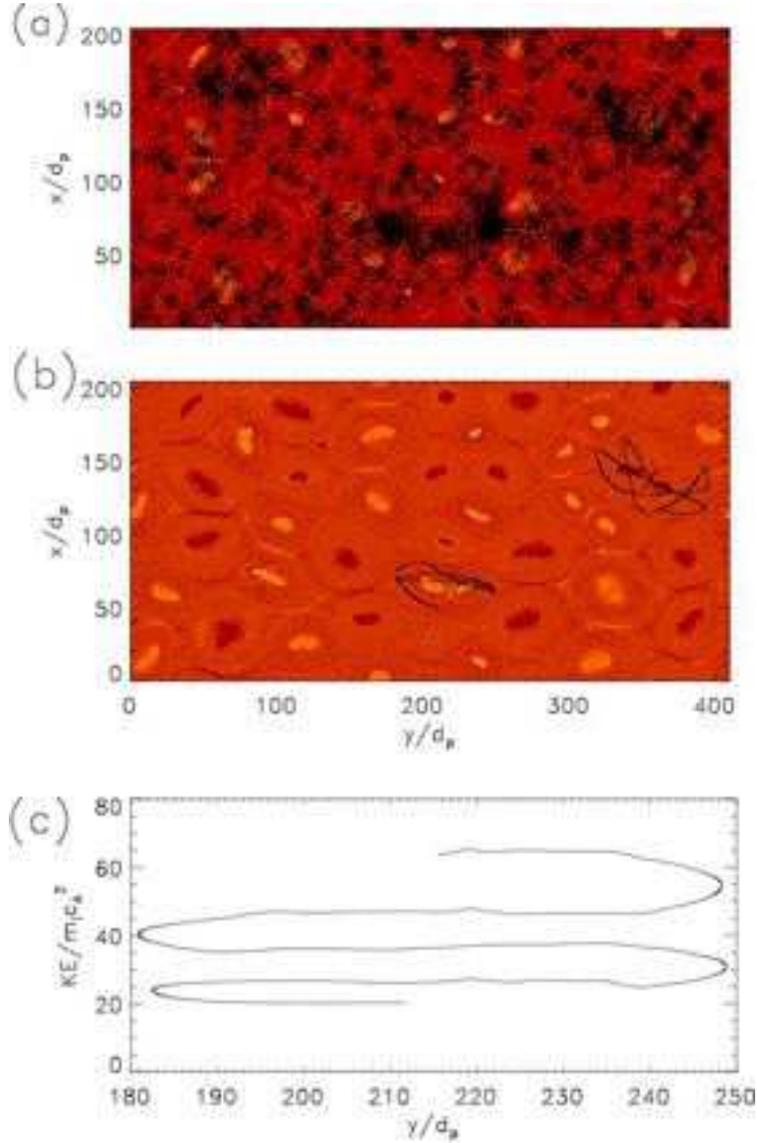}
\caption{\label{energy_traj} (Color online) In (a) the positions of
ions with energy exceeding $188m_ic_A^2$ at $\Omega_pt=150$ in a
background of $J_{ez}$ from the simulation in Fig.~\ref{jez3t}. In (b)
the orbits of test protons in the simulation fields at
$\Omega_pt=136$, during the time of most rapid energy gain of the seed
ACR particles, plotted over a background of $J_{ez}$. The particles
shown gained the most energy of a group of protons seeded with a
thermal spread equal to the ACR seed particles and randomly placed
near the two merging sites that contain the most energetic protons in
(a). In (c) the energy gain versus horizontal position of the proton
orbiting around $y\sim 210d_p$ in (b). The energy gain occurs during
reflection from the contracting ends of the island, a first order
Fermi process.}
\end{figure}

Anomalous cosmic rays are accelerated from LISM neutral particles that
are ionized deep in the heliosphere, picked-up by the supersonic solar
wind and carried out to the TS and heliosheath. The temperature of
these ions is around $1keV$ upstream of the TS and the pickup ions are
probably heated to $10keV$ at the TS although neither Voyager is able
to directly measure all of these high temperature ions
\citep{Decker08,Richardson08}. The background ions have a temperature
of around $10eV$ downstream of the shock. As discussed previously, to
explore how high temperature ions are accelerated during reconnection,
we carried out simulations with $5\%$ of the ions having an initial
temperature $30$ times that of the background. This fractional number
density is less than the realistic value of around $30\%$. The actual
pickup temperature is closer to $1000$ times that of the background
but the value used in the simulation is an upper limit based on the
requirement that these particles remain magnetized in our limited
computational domain.  In Fig.~\ref{energy_eiacr}(c) we show the
initial and final ($\Omega_pt=200$) ion energy spectra. The energy
distribution at late time maintains an exponential form. In
Fig.~\ref{energy_traj}(a) the locations of all ions with energies
above $188m_ic_A^2$ at $\Omega_pt=150$ are shown superimposed on a
background of the electron current $J_{ez}$. The most energetic ions
are clustered around three locations: $x,y=160d_p,180d_p$;
$x,y=60d_p,210d_p$; and $x,y=140d_p,360d_p$. Each of these locations
is within a large island where two islands had at an earlier time
merged. The merging process at these sites is evident in the the
out-of-plane current $J_{ez}$ in Fig.~\ref{jez3t}(b) and the ion
temperature data in Fig.~\ref{iontemp} at $\Omega_pt=136$, which is
the time at which the rate of energy gain of the ACR seed particles
maximizes. 

Detailed studies of the orbits and energy gain of test protons have
been carried out using the electric and magnetic fields from the
simulation at $\Omega_pt=136$, corresponding to
Fig.~\ref{jez3t}(b). The particles were initialized with thermal
distributions corresponding to $T_p=12.5m_pc_A^2$, the energy of the
seed ACRs, and randomly distributed in the vicinty of merging
islands. In Fig.~\ref{energy_traj}(b) we show two sample trajectories,
corresponding to particles which gained the most energy around their
respective merging sites. More details of the merging sites can be
seen in Fig.~\ref{jez3t}(b), where the orbits do not obscure the
structure of the current layer separating the merging islands. The
particles with the greatest energy gain bounce repetitively between
the contracting ends of the islands and gain energy during each
reflection. The energy gain of the particle orbiting around $y\sim
210d_p$ in Fig.~\ref{energy_traj}(b) is shown versus horizontal
position $y$ in Fig.~\ref{energy_traj}(c). The sharp increase in the
energy during each reflection is evident. The energy gain of this
proton exhibits the classic signatures of a first order Fermi process.
Thus, we conclude that the merging of two islands causes a contraction
of the resulting larger-scale island, which boosts the energy of ions
circulating within the larger island. Thus, the island contraction
mechanism seems to be responsible for the energy gain of the most
energetic ions in the simulation. Ions also gain energy within the
outflow jets of each of the current layers, as shown in
Fig.~\ref{iontemp} and as discussed earlier \citep{Drake09}.

The basic physics of particle acceleration in contracting islands,
which was seen in early test particle simulations \citep{Kliem94}, was
first discussed in detail in the context of electron acceleration
\citep{Drake06}. In Appendix A the basic concepts are reviewed. As
islands contract, particles that circulate around islands sufficiently
fast conserve their parallel (to ${\bf B}$) action $v_\parallel L$,
with $v_\parallel$ the parallel velocity and $L$ the length of the
field line. Island contraction shortens $L$ and therefore increases
$v_\parallel$. However, the corresponding reduction of the magnetic
field $B$ during contraction causes the perpendicular velocity
$v_\perp$ to decrease (due to the conservation of the magnetic
moment). During infinitesimal contractions in an isotropic plasma the
two effects cancel and there is no net energy gain. This result is
consistent with Parker's transport equation \citep{Parker65}, which
was derived in the limit of strong scattering, where the energy gain
is linked to plasma compression. Since 2-D island contraction preserves
the area within the island \citep{Fermo09} and is therefore nearly
incompressible, the Parker equation has no energy gain term
corresponding to island contraction. However, as shown in Appendix A
when an island undergoes a finite contraction, the parallel energy
gain dominates and as a result the rapidly circulating particles gain
net energy. This result, of course, requires that there is no
mechanism for scattering particles that is sufficiently strong to
maintain the isotropy of the plasma pressure as island contraction
takes place.

\begin{figure}
\epsscale{.6}
\plotone{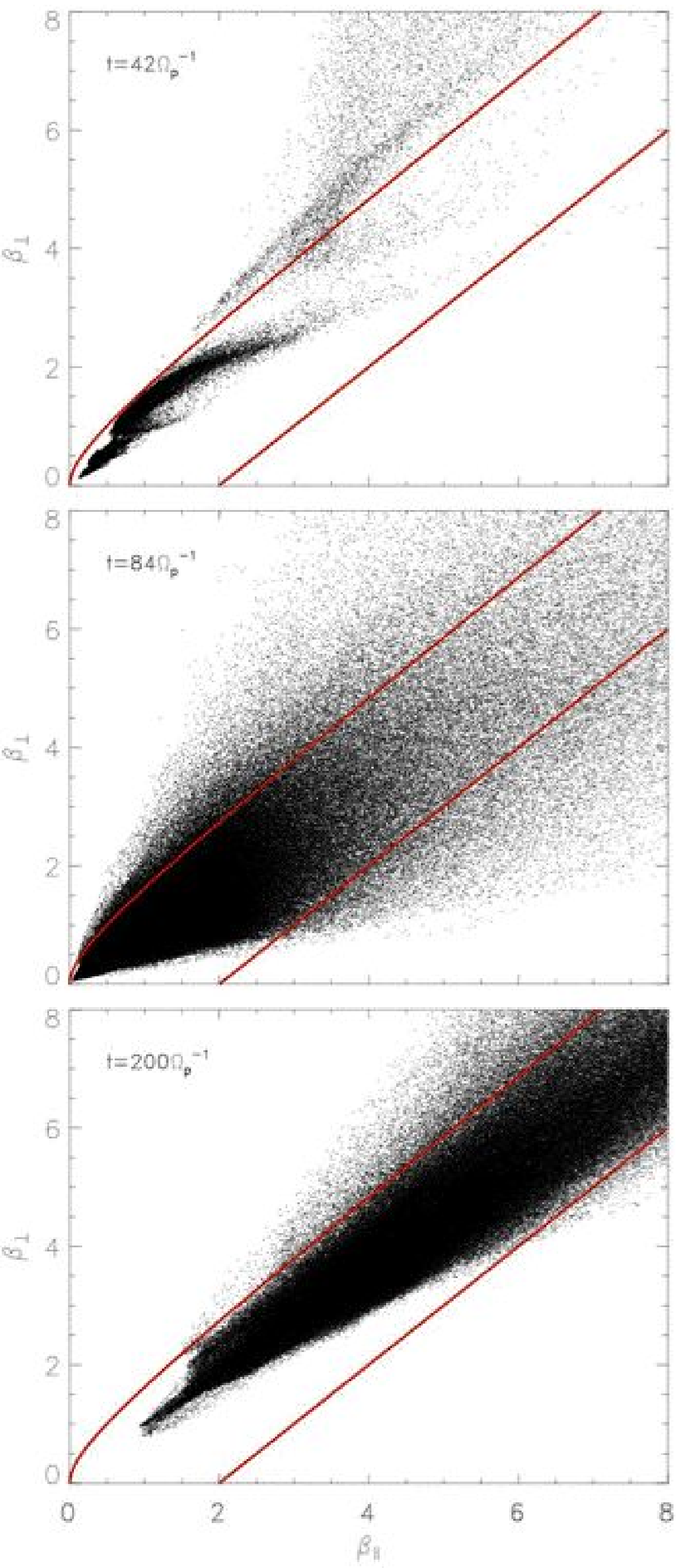}
\caption{\label{betaparperp} (Color online) Shown at three times in
the simulation of Fig.~\ref{energy_eiacr}(c) (no ACR seed particles)
are the values of $\beta_\parallel$ and $\beta_\perp$ with each grid
point represented by a point. The upper and lower lines in each plot
are the marginal stability conditions for the mirror and firehose
modes, respectively. The band between the two curves is the nominal
stable region.}
\end{figure}

The spatial differences between the parallel and perpendicular
temperatures in Fig.~\ref{iontemp} are evidence that if some mechanism
does cause particle scattering in the simulations, it is sufficiently
weak to allow significant pressure anisotropies to develop during
island growth and merger. To put this on a more quantitative basis, in
Fig.~\ref{betaparperp} we show data showing the distribution of
$\beta_\parallel=8\pi p_\parallel/B^2$ and $\beta_\perp=8\pi
p_\perp/B^2$ at three times from the simulation of
Fig.~\ref{energy_eiacr}(a,b) (no ACR seed particles). Each point is
from a single grid point of the simulation. The two curves are the
marginal stability boundaries for the firehose,
\begin{equation}
\beta_\parallel=\beta_\perp+2,
\label{fh}
\end{equation}
and mirror instabilities,
\begin{equation}
\beta_\parallel=\frac{\beta_\perp^2}{1+\beta_\perp}.
\label{mirr}
\end{equation}
Points below the lower curves are in the region of firehose
instability while those above the upper curves are in the region of
mirror mode instability. Both instabilities act to isotropize the
plasma in their respective regions of instability. We emphasize, of
course, that these stability boundaries are based on simple
homogeneous models and do not reflect the complex geometry of
reconnecting islands. At early time in Fig.~\ref{betaparperp}(a) the
bulk of the plasma is nearly isotropic with a few scattered points
lying in the mirror unstable region. At $\Omega_pt=84$ in
Fig.~\ref{betaparperp}(b), which is a time of strong reconnection and
particle energy gain in this simulation, significant anisotropy has
developed and the data has spilled into the unstable regions of both
the mirror and firehose instabilities. At late time in
Fig.~\ref{betaparperp}(c) the plasma has largely retreated into the
stable region between the two stability boundaries. The firehose
condition in particular appears to significantly constrain the
increase in $\beta_\parallel$. Although plotted in a slightly
different format, the data in Fig.~\ref{betaparperp}(c) bears a
striking similarity to data from the slow solar wind
\citep{Hellinger06,Bale09}. In the data from the solar wind, as in the
present simulation, the plasma bumps up against the theoretical
stability boundaries of the mirror and oblique firehose stability
boundaries. Moreover, in the solar wind enhanced magnetic fluctuations
are measured close to the stability boundaries, suggesting that the
two instabilities are limiting the plasma anisotropy
\citep{Bale09}. The overall increase in the number of points at higher
values of $\beta_\parallel$ and $\beta_\perp$ with increasing time in
Fig.~\ref{betaparperp} reflects the strong plasma heating and the
reduction in the magnetic energy as the simulation develops. This data
demonstrates that, while the plasma dynamics act to limit the level
of anisotropy, significant anisotropies do develop, which, as shown in
the Apprendix, is a requirement for net energy gain to take place
during island contraction.

\begin{figure}
\epsscale{.6}
\plotone{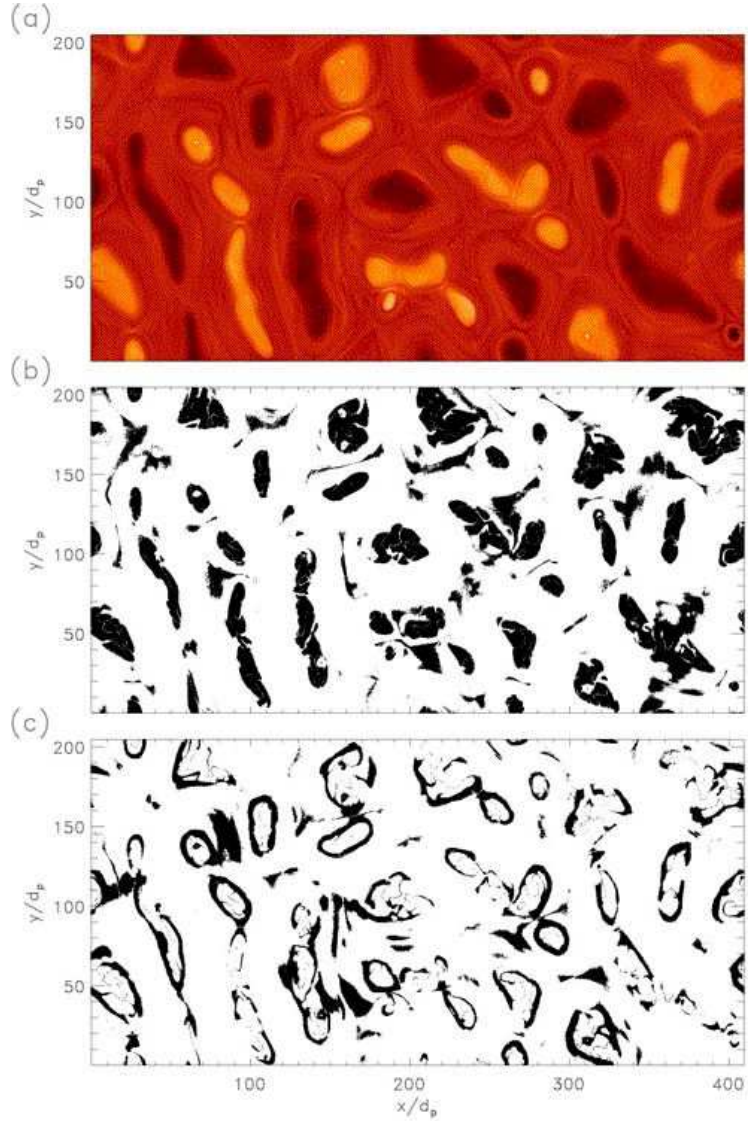}
\caption{\label{fhmirr} (Color online) Data at $\Omega_pt=200$ from the simulation in Fig.~\ref{betaparperp}. In (a) the out-of-plane current $J_z$ showing the structure of magnetic islands at this time. In (b) and (c) the black denotes regions where the firehose and mirror mode marginal stability conditions are, respectively, violated.}
\end{figure}

It has been suggested previously that the increase in parallel
electron energy through island contraction would be limited by the
firehose instability \citep{Drake06}. In an anisotropic plasma the
tension force ${\bf F}_t$ of bent magnetic field fields takes the
form,
\begin{equation}
{\bf F}_t=(1-\frac{\beta_\parallel-\beta_\perp}{2}){\bf B\cdot\nabla B}.
\end{equation}
As the firehose marginal condition is approached the tension force
goes to zero and island contraction stops, halting reconnection and
the associated accelereration of particles. In the Appendix this
behavior is shown in a simple model of a contracting flux loop. In
Fig.~\ref{fhmirr} we present data from $\Omega_pt=200$, corresponding
to Fig.~\ref{betaparperp}(c), on the spatial distribution of regions
where the firehose and mirror stability boundaries are violated. In
Fig.~\ref{fhmirr}(a) is the out-of-plane current, which shows that
structure of islands at this time. In Fig.~\ref{fhmirr}(b,c) the black
marks the regions where the firehose and mirror conditions are
violated, respectively. As expected from Fig.~\ref{iontemp}, the
islands are regions where the firehose condition is violated while the
x-line regions and separatrices are where the mirror condition is
violated. The firehose condition in Fig.~\ref{fhmirr}(b) combined with
the results of Fig.~\ref{betaparperp}(c) strongly suggest that in our
simulations the firehose condition does play a central role limiting
island contraction during reconnection, consistent with the model
presented in the Appendix.

Further evidence that the island contraction mechanism plays the
dominant role in heating the highest energy particles can be gleaned
from the energy spectra in Fig.~\ref{energy_eiacr}(c). A model equation for the the omnidirection particle distribution function $F(v,t)=4\pi v^2f(v,t)$ in a homogeneous system is given by
\begin{equation}
\frac{\partial F}{\partial t}+\frac{\partial}{\partial v}\dot{v}F=-\frac{F}{\tau},
\label{Fequation}
\end{equation}
where $\dot{v}$ is the rate of increase in the particle speed and
$\tau$ is the loss rate of particles from the acceleration region. For the simulations there is no particle loss so $\tau\rightarrow\infty$ and the change in $F$ during a short time interval $\delta t$ is given by
\begin{equation}
\delta F\simeq -\delta t\frac{\partial}{\partial v}\dot{v}F\simeq\frac{\dot{E}\delta t}{T_i}F,
\end{equation}
where $E=mv^2/2$ is the energy, we have taken $F$ to be a Maxwellian with a temperature $T_i$ and for $E\gg T_i$ the velocity derivative acts only on $F$. Thus,\begin{equation}
\delta(\ln F)\simeq \frac{\dot{E}\delta t}{T_i}.
\end{equation}
Consistency with Fig.~\ref{energy_eiacr}(c), in which the increment of
$\ln F$, increases linearly with $E$, requires that $\dot{E}\propto
E$, which suggests that the energy gain is through a first order
Fermi process. 

In the island contraction mechanism the average rate of
energy gain is proportional to $<c_A/L>$, where $c_A$ is the Alfv\'en
speed based on the magnetic field of the island before contraction and
$L$ is the associated island length. The brackets denote an average
over the islands in the system. We estimate this average for the
simulation of Fig.~\ref{iontemp}(c) at $\Omega_pt=136$ by computing
the average of $|({\bf B}\cdot{\bf\nabla B})/B|$. The result is
$0.05B_0/d_p$, which corresponds to $L\sim 20d_p$. This scale is
consistent with the size of the islands in
Fig.~\ref{iontemp}. From the rate of change of $\ln F$ during the time
interval $\Omega_pt=100-150$ we calculate the rate of ion energy
gain,
\begin{equation}
\dot{E}=0.05\langle\frac{c_A}{L}\rangle E.
\label{Edot}
\end{equation}
The factor $0.05$ represents the weighted area where islands are
undergoing contraction.

\section{ACR spectra resulting from reconnection}
The reconnection of the sectored magnetic field dissipates most of the
magnetic free energy. The simulations presented in Sec.~\ref{picsims}
demonstrated that reconnection of the sectored field accelerates
seed ACRs through the magnetic island contraction mechanism, which is
a first order Fermi process. Furthermore island contraction is regulated by the approach to the firehose marginal stability condition (Figs.~\ref{betaparperp} and \ref{fhmirr}). However, the PIC simulations can not
produce the expected ACR energy spectrum because the islands are
unrealistically small, and therefore do not have enough flux to contain
$100MeV/$nucleon ions, and because a loss mechanism is needed to balance the
source in the energy drive equation to obtain a steady state
spectrum. 

Knowing the functional form of the rate of energy gain of energetic
ions from the simulations, however, we propose a simple 1-D model
equation for the ACR energy spectrum that parallels the 2-D model
proposed earlier for electrons \citep{Drake06}. The basic idea is that
the ions that gain significant energy undergo acceleration in a series
of contracting islands. As discussed earlier, the energy gain of an
ion undergoing Fermi acceleration in a single island is controlled by
the invariance of the parallel action. The action invariant limits the
ion energy gain in a single contraction and requires particles
undergoing significant energy gain to interact with many islands
\citep{Drake06}. What was also shown earlier \citep{Drake06} was that particles
undergoing acceleration slowly drift outwards in the island and
eventually cross a narrow boundary layer at the separatrix bounding
the island. During the crossing of the boundary layer, which has a
scale length controlled by electrons, the pitch angle of the particles
is scattered. Because the Larmor radius of ions is greater than that
of electrons, the ions are even more effectively scattered -- the
trajectories of the most energetic particles in Fig.~\ref{energy_traj}
are only marginally magnetized. Thus, the particles will undergo
acceleration in a series of contracting islands, starting with an
approximately isotropic distribution at the beginning of each
contraction. This perhaps understates the rate of ion acceleration
since it is possible that $p_\parallel$ may be greater than $p_\perp$
as contraction initiates. The rate of gain of energy of ions of a
given energy can be calculated by averaging over the rate of energy
gain over the distribution of islands in the simulation as shown in
Eq.~(\ref{rate}). If particle acceleration begins in a system in which
the plasma $\beta$ is initially very low, we show in the Appendix
that, as expected, the contraction velocity $u$ of the island is the Alfv\'en
speed. As time passes the energetic particle pressure builds up. When
the initial normalized particle pressure $\beta$ approaches unity, the
island can only contract a small amount until the marginal firehose
condition is reached and the tension force that drives reconnection
goes to zero, contraction stops and particle energy gain ceases. This
was demonstrated explicitly in the case of electrons
\citep{Drake06}. The violation of the marginal firehose condition
within islands in Fig.~\ref{fhmirr} demonstrates that the same
important feedback on the islands is taking place at the ion
scales. That feedback is a critical element of any reconnection driven
particle acceleration model was an important conclusion of an MHD
study of reconnection that included the dynamics of test particles
\citep{Onofri06}.

To model the shutoff of particle acceleration as the plasma $\beta$
rises (see Eq.~(\ref{uhighbeta})), we take a simple form for the
contraction velocity,
\begin{equation}
u=c_{A}\left(1-\frac{4\pi p}{B^2}\right)^{1/2}
\label{u}
\end{equation}
with $p$ the ACR pressure. Averaging over the distribution of islands in the simulation we obtain the averaged rate of ion energy gain 
\begin{equation}
\frac{dE}{dt}=0.05\langle\frac{c_A}{L}\rangle\left(1-\frac{4\pi
p}{B^2}\right)^{1/2}E,
\label{Edotfh}
\end{equation}
which differs from the result benchmarked with the simulation data in
Eq.~(\ref{Edot}) only by the factor within the parenthesis reflecting
inability of islands to contract as $\beta$ approaches unity. To
facilitate a direct evaluation of the energy spectra, we consider a
simple 1-D model for the omnidirectional distribution function
$F(v,t)$ in which convective loss is simply represented as a loss time
$\tau_L$. Using the result in Eq.~(\ref{Edotfh}) to calculate the rate
of convection $\dot{v}$ in phase space, we obtain the equation for
$F$,
\begin{equation}
\frac{\partial F}{\partial t}+\frac{1}{\tau_h}\left(1-\frac{4\pi
p}{B^2}\right)^{1/2}\frac{\partial}{\partial
v}vF=-\frac{1}{\tau_L}(F-F_0);
\label{fpequation}
\end{equation}
where the ACR pressure is given by 
\begin{equation}
p=\int_0^\infty dv F(v,t)mv^2,
\label{p}
\end{equation}
$1/\tau_h=0.025\langle c_A/L\rangle$ is the pickup ion heating
rate and $F_0$ is the ACR seed. This equation differs from that
presented in \cite{Drake06} because of the absence of magnetic shear
in the sectored heliospheric magnetic field and the 1-D assumption.

The particle loss represented by $\tau_L$ limits the energy gain of
the ACRs. Since the ions are trapped by the magnetic islands, the loss
rate is controlled by the large-scale convective flows in the
HS shown in Fig.~\ref{mhdB}. Thus, $\tau_L$ is energy
independent and is of the order of the heliospheric covection time,
which can become very long near the stagnation point of the flow
near the HP. The heating time is the Alfv\'en transit time
across a sector and, in contrast with the loss time, decreases near
the heliopause due to the increase of $c_A$ and the reduction of the
sector spacing (see Fig.~\ref{mhdB}). We therefore conclude that
$\tau_L\gg\tau_h$ in the sectored region of interest for ACR
acceleration.

Equation (\ref{fpequation}) is an equidimensional equation for
sufficiently high velocity where the source $F_0$ can be
neglected. The equation therefore has no intrinsic velocity scale and
the solutions in this region are therefore power laws. Fortunately, the
steady state solution to Eq.~(\ref{fpequation}) can be written in
closed form,
\begin{equation}
F(v)=(\gamma -1)v^{-\gamma}\int_0^\infty dss^{\gamma -1}F_0(s),
\label{F}
\end{equation}
where the power law $\gamma$ satisfies the equation,
\begin{equation}
(\gamma -1)\left(1-\frac{4\pi p_0}{B^2}\frac{\gamma -1}{\gamma -3}\right)^{1/2}=\frac{\tau_h}{\tau_L},
\label{gamma}
\end{equation}
and the ACR pressure was evaluated exactly $p=p_0(\gamma -1)/(\gamma
-3)$ with $p_0$ the initial pickup particle pressure. The convergence
of the ACR pressure requires $\gamma > 3$. We also emphasize that the
expression for $\gamma$ in Eq.~(\ref{gamma}) is valid only if the seed ACR
$\beta$ is small, which is not generally valid within the heliosheath but
as shown in Fig.~\ref{mhdB}(b) is valid close to the heliopause. It is
only in this low $\beta$ region that the magnetic field has sufficient
energy to raise the pickup seed particles to energies typical of the
ACRs. Equation (\ref{gamma}) has simple solutions in the limit of
large and small $\tau_h/\tau_L$ as follows:
\begin{equation}
\gamma = \frac{\tau_h}{\tau_L} \qquad \tau_h/\tau_L\gg 1
\end{equation}
\begin{equation}
\gamma=3+\beta_0  \qquad \tau_h/\tau_L\ll 1,
\end{equation} 
where $\beta_0=8\pi p_0/B^2$ is the initial pickup $\beta$.

 \section{DISCUSSION AND IMPLICATIONS}
\label{discussion}

The energy spectra of the ACRs have continued to unroll as the Voyager
spacecraft have moved further into the HS and there is evidence that
the source region of the higher energy particles has now been reached
\citep{Stone08}. The spectral indices of the ACR H and He spectra are
around $1.75$. In the reconnection model presented here the power law
index of the differential energy spectrum is $\gamma/2=(3+\beta_0)/2$.
The MHD data shown in Fig.~\ref{mhdB}(b) indicate that close to the
HP the plasma $\beta$ drops to $0.5$ so the estimated spectral index
of the reconnection model approaches $1.75$, which is close to the
value observed.

A central question, of course, is whether the rate of energy gain due
to reconnection is sufficient to produce ACRs in the range of
$100MeV/nuc$. The heating rate in Eq.~(\ref{Edot}) depends on the typical
Alfv\'en transit time across a magnetic island. The characteristic
scale length of the magnetic islands is the sector spacing. An upper
limit on the sector spacing is given by the $13$ day sector periodicity
times the local heliosheath velocity, which is around $100km/s$
\citep{Richardson08}. Based on a local magnetic field of $0.15nT$
\citep{Burlaga08} and a density of $0.002/cm^3$ \citep{Richardson08},
we obtain an Alfv\'en velocity of $74km/s$. The Alfv\'en transit time
across a typical magnetic island is therefore $18$ days. From
Eq.~(\ref{Edot}) the characteristic heating time $\tau_h$ is around a
year. The compression of the magnetic field and reduction of the
density near the HP can increase $c_A$ by a factor of three, which
reduces the heating time to $120$ days. The total e-folding time to
increase the $10keV/nuc$ pickup ions to $100MeV/nuc$ therefore ranges from $3$
to $9$ years. The convective flow out of the sector field in the
latitudinal direction very likely dominates the loss of energetic
particles. For a sector-zone spanning $30^\circ$ the characteristic
width of the sector zone at a HP distance of $150AU$ is around $75AU$. A
transverse velocity of $30km/s$ or $6AU/$year yields $\tau_L=13$ years,
which exceeds $\tau_h$ as expected. 

A key observation of ACRs is the similarity in the spectra of
different species when expressed on a per nucleon basis. In the case
of the shock acceleration model this result is because the rate of
energy gain of a particle as it reflects back and forth across the
shock is on average independent of mass when expressed on a per
nucleon basis \citep{Blandford87}. The contracting island mechanism
for particle energy gain is analagous -- upon reflection from the end
of a contracting island, the rate of energy gain is independent of
mass when expressed on a per nucleon basis. In particular, the
particle distribution functions of the various minor species are also
given by the result in Eq.~(\ref{F}), where $F_0$ is the distribution
of seed particles of a given species but the spectral index $\gamma$
is controlled by H with some contribution from He. Thus, at high
energy the spectra of all of the minor species take the form of
power laws with the same spectral indices as H.

A surprising observation is the nearly universal $f\propto v^{-5}$
spectrum of super-Alfv\'enic ions observed in the quiet-time solar
wind throughout the heliosphere \citep{Fisk06}. These distributions
correspond to $F\propto v^{-3}$ or $F\propto E^{-1.5}$, the low
$\beta$ limit of Eq.~(\ref{gamma}). An explanation of this universal
spectral index has been offered based on ion acceleration in
compressible turbulence \citep{Fisk06}. We suggest that the non-pickup
ions accelerated through reconnection within the heliosheath could be
the source of this near universal spectrum of super-Alfv\'enic
particles. Since the pressure of the non-pickup particles is much less
than that of the pickup particles, they can be treated as minor
species just like the minor ACRs. Since they start from lower energy
they never reach energies associated with ACRs, but like the minor
species their spectral index is controlled by that of the pickup
protons. In the limit of low $\beta_0$, of course, the spectral index
is $1.5$. The spectral index of $1.5$ arises from the approach to the
marginal firehose condition of the ACR protons. An important question
is whether such a spectrum of super-Alfv\'enic ions can move upstream
across the TS and into the inner heliosphere. An exploration of the
transport of such particles is beyond the scope of the present
manuscript but should be pursued.

Finally, we note that the reconnection of the sectored magnetic field
in the heliosheath was independently proposed as the source of the
ACRs by \cite{Lazarian09}, denoted by LO in the following. Beyond the
basic idea that the sectored field is a source for ACR acceleration
and that the compression of the sectors on the approach to the
heliopause would enhance the likelihood that reconnection would take
place, the LO model is very different from that discussed
here. Reconnection in the LO model is based on the turbulent
reconnection model of \cite{Lazarian99} although the explicit source
of the turbulence is not identified. In the model presented in this
manuscript reconnection is collisionless and fast reconnection is
facilitated by the Hall term in Ohm's law \citep{Birn01}. Recent
scaling studies \citep{Shay07} and observations \citep{Phan07} confirm
that Hall reconnection remains fast even in systems comparable in size
to that defined by the sector spacing close to the HP. Further, in the
model presented here the interaction of islands on adjacent current
layers facilitates the rapid dissipation of the magnetic energy in the
sectored region. While the mechanism for particle acceleration in both
models is a first order Fermi process, the LO model is based on
periodic reflection from the plasma flowing inward toward the x-line
\citep{Lazarian05} and curiously is taken in the strongly relativistic
limit and so does not apply to ACRs. There is no evidence in even the
largest scale PIC simulations carried out to date that such a
mechanism is active during reconnection (see the particle
distributions in Fig.~2 of \cite{Drake09}). On the other hand it can't
be ruled out that some additional source of turbulence and associated
scattering could alter the particle dynamics allowing particle
acceleration as proposed by \cite{Lazarian05} to take place. One might
make the case that the particles moving along the trajectories shown
in Fig.~\ref{energy_traj} are reflecting from the inflow of the
reconnecting islands and the acceleration mechanism is therefore
similar to that of the LO model. However, the spectra resulting from
this acceleration process are not given by the \cite{Lazarian05}
model.

\acknowledgments
This work has been supported by NSF Grant PHY-0316197 and NASA Grant
NNG06GH23G.  M.\ O.\ acknowledges the support of a NASA-Voyager
Guest Investigator grant NNX07AH20G and an NSF CAREER Grant
ATM-0747654. Computations were carried out at the
National Energy Research Scientific Computing Center and the NASA Ames
Research Center.

\appendix
\section{Particle acceleration during island contraction}
The magnetic energy that is released during magnetic reconnection
takes place as bent magnetic fields downstream from the x-line
straighten out, driving the outflow exhaust. In the Petschek model the
bent fields take the form of slow shocks that both drive the outflow
and heat the plasma. During reconnection in a multi-island
environment, the contraction of magnetic islands releases magnetic
energy -- initially oblate islands try to become round
\citep{Drake06}. The merging and contraction of islands is a nearly
incompressible process \citep{Fermo09}. In magnetic reconnection,
therefore, particle acceleration is through incompressible flows in
contrast with shocks, where particle energy gain as described by Parker's
equation is explicitly driven by plasma compression. 

A fundamental question is therefore whether the incompressible flows
which dominate the dynamics of islands during reconnection can
accelerate particles through a first-order Fermi process or it is
necessary to identify another mechanism for particle acceleration. If
a first order Fermi mechanism does take place in the presence of
incompressible flows, how do we resolve the apparent contradiction
with the Parker equation and the conventional wisdom that first-order
Fermi acceleration requires compressible plasma motion? 

To make analytic progress, we consider a simple 2-D elongated ring of
magnetic flux in the shape of a racetrack with an initial length along
a field line $L_0$ and an initial width (of the track) $w_0$ such that
$L_0 \gg w_0$. The magnetic field in the flux tube is taken to be
$B_0$. The flux tube is not in equilibrium, which is not an issue
because we are interested in test particle behavior as the ring
contracts. During contraction, the flux $\psi_0=Bw$ is conserved as is
the total area $A_0=Lw$ of the flux tube so the magnetic field $B$ can
be calculated as a function of its initial value as $B=B_0L/L_0$. The
constants of the particle motion are the magnetic moment
$\mu=mv_\perp^2/B$ and the longitudinal action $J_\parallel
=v_\parallel L$. These invariants allow us to calculate the particle
velocity $v$ in terms of the initial velocities and $L$,
\begin{equation}
v^2=v_{\perp 0}^2\frac{L}{L_0}+v_{\parallel 0}^2\frac{L_0^2}{L^2}.
\end{equation}
If we linearize this result by assuming that $L$ changes by only a
small amount $\Delta L$, we find the change in $v^2$,
\begin{equation}
\Delta v^2=(v_{\perp 0}^2-2v_{\parallel 0}^2)\frac{\Delta L}{L_0}.
\end{equation}
This result is equivalent to that obtained by \citep{Cho06} in their
evaluation of particle acceleration in incompressible MHD turbulence
if their parallel flow velocity is equated with the rate of change
of the field line length. For an initially isotropic plasma $v_{\perp
0}^2=2v_{\parallel 0}^2$ (in an averaged sense) so the energy gain is
zero. This is because the increase in the parallel energy
(conservation of $J_\parallel$) is balanced by a reduction of the
perpendicular energy (conservation of $\mu$ and the decrease in
$B$). This result is consistent with the Parker equation, in which
there is no energy gain for incompressible flows since the Parker
equation was derived under the assumption of strong scattering. On the
other hand, even if the plasma is initially isotropic a large
contraction or stretching of the island increases the particle energy,
\begin{equation}
v^2=v_0^2(\frac{2L}{3L_0}+\frac{L_0^2}{3L^2}),
\label{vsq}
\end{equation}
assuming that there is no scattering during contraction. Stretching is
magnetically unfavorable and is therefore of less interest during
reconnection and island merger. The result in Eq. (\ref{vsq}) is
consistent with the PIC simulations of island contraction
\citep{Drake06}, which started with an initial elongated island in a
nearly isotropic plasma but nevertheless revealed strong parallel
heating of electrons and substantial net energy gain. To the extent
that nearly collisionless plasma can maintain anisotropy, with limits
perhaps controlled by the firehose or a related marginal stability
condition, contracting islands drive particle acceleration through a
first-order Fermi process.

To gain further insight into how the full distribution of particles
gains energy during the contraction of an island and the role of the
firehose condition, we consider an elongated island of length $L_0$
with an initially isotropic distribution of particles with initial
pressure $p_0$. The parallel and perpendicular pressures during
contraction are given by
\begin{equation}
p_\parallel=p_0\frac{L_0^2}{L^2},\qquad p_\perp=p_0\frac{L}{L_0}. 
\end{equation}
Writing an energy equation for the contracting loop, including the
internal energy $(p_\parallel+2p_\perp)/2$, the magnetic energy
$B^2/8\pi$ and the flow energy associated with the contraction
velocity $u$, $\rho u^2/2$, we find
\begin{equation}
\frac{u^2}{c_{A0}^2}+\left(\frac{L^2}{L_0^2}-1\right) +\frac{\beta_0}{2}\left( \frac{L_0^2}{L^2}+2\frac{L}{L_0}-3\right) =0,
\label{energy}
\end{equation}
where $c_{A0}^2=B_0^2/4\pi\rho$ and $\beta_0=8\pi p_0/B_0^2$. In
Eq.~(\ref{energy}) the first term is the flow energy, the second the
change in magnetic energy and the third the change in internal
energy. In writing this equation we have ignored geometrical factors
of order unity that enter the flow energy due to the fact that the
convective velocity varies over the length of the island. In the
absence of the pressure (the limit of low $\beta_0$), the reduction of
the magnetic energy as $L$ decreases well below $L_0$ causes $u$ to
increase up to the Alfv\'en speed. However, if $\beta_0$ is not too
small, the increase in the total internal energy can balance the
reduction in magnetic energy so that $u$ takes on a maximum
value. Taking the differential of the second two terms with respect to
$L$ yields an equation for the value of $L$ at the peak velocity;
\begin{equation}
2\frac{L}{L_0}+\beta_0\left( 1-\frac{L_0^3}{L^3}\right) =0.
\end{equation}
This condition is precisely the marginal firehose condition
\begin{equation}
p_\parallel-p_\perp-\frac{B^2}{4\pi}=0.
\end{equation} 
Thus, the rate of island contraction continues to increase with
$p_\parallel$ increasing compared with $p_\perp$ until the marginal
firehose condition is violated at which point the net force driving
loop contraction drops to zero. Since loop contraction is the driver
of reconnection, this causes reconnection to cease. 

It is illuminating to compare the maximum contraction velocity $u$ in
the low and high $\beta_0$ limits. At low $\beta_0$ the maximum
velocity occurs for $L/L_0=(\beta_0 /2)^{1/4}$ and 
\begin{equation} 
u\simeq c_{A0},\qquad \beta_0 \ll 1,
\label{ulowbeta}
\end{equation}
while for $\beta_0$ large $L/L_0\simeq 1-2/(3\beta_0)$ and 

\begin{equation}
u\simeq\frac{2c_{A0}}{\sqrt{3\beta_0}},\qquad \beta_0 \gg 1. 
\label{uhighbeta}
\end{equation}
Thus, at high $\beta_0$ only a very small contraction of the
island takes place and the contraction velocity and associated
particle acceleration are weak.

\clearpage

\end{document}